\documentclass[prb,twocolumn,showpacs]{revtex4}
\pdfoutput=1
\usepackage{amsfonts} 
\usepackage{amsmath}
\usepackage{amssymb}
\usepackage{bm}
\usepackage[usenames, dvipsnames]{color} 
\usepackage{dcolumn}
\usepackage{epic} 
\usepackage{epsfig}
\usepackage{graphicx}

\usepackage[breaklinks,colorlinks = true,linkcolor = blue,urlcolor=blue,citecolor=blue]{hyperref}
\usepackage{soul}
\usepackage{subfigure}
\usepackage{times}
\usepackage{ulem}
\usepackage{wrapfig} 
\usepackage{xy} 

\newcommand{\beq}{\begin{equation}}
\newcommand{\eeq}{\end{equation}}
\newcommand{\beqa}{\begin{eqnarray}}
\newcommand{\eeqa}{\end{eqnarray}}

\begin{document}

\title{Spin-orbital liquids in non-Kramers magnet on Kagome lattice}
\author{Robert Schaffer$^1$}
\author{Subhro Bhattacharjee$^{1,2}$}
\author{Yong Baek Kim$^{1,3}$}
\affiliation{$^1$ Department of Physics and Center for Quantum Materials, University of Toronto, Toronto, Ontario M5S 1A7, Canada.\\
$^2$ Department of Physics and Astronomy, McMaster University, Hamilton, Ontario  L8S 4M1, Canada.\\
$^3$ School of Physics, Korea Institute for Advanced Study, Seoul 130-722, Korea.}

\date{\today}
\begin{abstract}
Localized magnetic moments with crystal-field doublet or pseudo-spin 1/2
may arise in correlated insulators with even number of electrons and strong
spin-orbit coupling. Such a non-Kramers pseudo-spin 1/2 is the consequence
of crystalline symmetries as opposed to the Kramers doublet arising from
time-reversal invariance, and is necessarily a composite of spin and orbital
degrees of freedom. We investigate possible spin-orbital liquids with fermionic
spinons for such non-Kramers pseudo-spin 1/2 systems on the Kagome lattice.
Using the projective symmetry group analysis, we find {\it ten} new
phases that are not allowed in the corresponding Kramers systems.
These new phases are allowed due to unusual action of the time reversal operation
on non-Kramers pseudo-spins. We compute the spin-spin dynamic structure factor that shows characteristic features of these non-Kramers spin-orbital liquids arising from their unusual coupling to neutrons, which is therefore relevant for neutron scattering experiments. We also point out possible anomalous broadening of Raman scattering intensity that may serve as a signature experimental feature for gapless non-Kramers spin-orbital liquids.
\end{abstract}

\maketitle
\section{Introduction}
The low energy magnetic degrees of freedom of a Mott insulator, in the presence of strong spin-orbit coupling, are described by states with entangled spin and orbital wave functions.\cite{1999_fazekas,2001_yosida} In certain crystalline materials, for ions with even numbers of electrons, a low energy spin-orbit entangled ``pseudo-spin"-1/2 may emerge, which is not protected by time-reversal symmetry (Kramers degeneracy)\cite{1930_kramers} but rather by the crystal symmetries.\cite{1952_bleany,2011_onada} Various phases of such {\it non-Kramers} pseudo-spin systems on geometrically frustrated lattices, particularly various quantum paramagnetic phases, are of much recent theoretical and experimental interest in the context of a number of rare earth materials including frustrated pyrochlores\cite{2002_matsuhira,2009_matsuhira,2006_nakatsuji,2013_flint,2012_sungbin} and heavy fermion systems.\cite{2013_chandra,2006_suzuki}

In this paper, we explore novel {\it spin-orbital liquids} that may emerge in these systems due to the unusual transformation of the non-Kramers pseudo-spins under the time reversal transformation. Contrary to Kramers spin-1/2, where the spins transform as ${\bf S}\rightarrow -{\bf S}$ under time reversal,\cite{1930_kramers} here only one component of the pseudo-spin operators changes sign under time reversal: $\{\sigma^1,\sigma^2,\sigma^3\}\rightarrow \{\sigma^1,\sigma^2,-\sigma^3\}$.\cite{1952_bleany,2011_onada} This is because, due to the nature of the wave-function content, the $\sigma_3$ component of the pseudo-spin carries a dipolar magnetic moment while the other two components carry quadrupolar moments of the underlying electrons. Hence the time reversal operator for the non-Kramers pseudo-spins is given by $\mathcal{T}=\sigma^1 K$ (where $K$ is the complex conjugation operator), which allows for new spin-orbital liquid phases. Since the magnetic degrees of freedom are composed out of wave functions with entangled spin and orbital components, we prefer to refer the above quantum paramagnetic states as spin-orbital liquids, rather than spin liquids.

Since the degeneracy of the non-Kramers doublet is protected by crystal symmetries, the transformation properties of the pseudo-spin under various lattice symmetries intimately depend on the content of the wave-functions that make up the doublet. To this end, we focus our attention on the example of Praseodymium ions (Pr$^{3+}$) in a local $D_{3d}$ environment, which is a well known non-Kramers ion that occurs in a number of materials with interesting properties.\cite{2002_matsuhira,2009_matsuhira,2006_nakatsuji} Such an environment typically occurs in Praseodymium pyrochlores given by the generic formulae Pr$_2$TM$_2$O$_7$, where TM(= Zr, Sn, Hf, or Ir) is a transition metal. In these compounds, the Pr$^{3+}$ ions host a pair of 4$f$ electrons which form a $J=4$ ground state manifold with $S=1$ and $L=5$, as expected due to Hund's rules. In terms of this local environment we have a nine fold degeneracy of the electronic states.\cite{2011_onada} This degeneracy is broken by the crystalline electric field. The oxygen and TM ions form a $D_{3d}$ local symmetry environment around the Pr$^{3+}$ ions, splitting the nine fold degeneracy. A standard analysis of the symmetries of this system (see appendix \ref{app:D3d}) shows that the $J=4$ manifold splits into three doublets and three singlets ($\Gamma_{j=4} = 3E_g +2A_{1g} + A_{2g}$) out of which one of the doublets is found to have the lowest energy, usually well separated from the other crystal field states.\cite{2011_onada} This doublet (details in Appendix \ref{app:D3d}), formed out of a linear combination of the $J_z = \pm 4$ with $J_z = \pm 1$ and $J_z = \pm 2$ states, is given by
\begin{align}
|\pm\rangle &= \alpha |m=\pm 4\rangle \pm \beta |m=\pm 1\rangle - \gamma|m=\mp2\rangle.
\label{eq_wavefn}
\end{align}
The non-Kramers nature of this doublet is evident from the nature of the ``spin" raising and lowering operators within the doublet manifold; the projection of the angular momentum raising and lowering operators to the space of doublets is zero ($\mathcal{P} J^{\pm}\mathcal{P}|\sigma\rangle = 0$ where $\mathcal{P}$ projects into the doublet manifold). However, the projection of the $J^z$ operator to this manifold is non-zero, and describes the z component of the pseudo-spin ($\sigma^3$). In addition, there is a non-trivial projection of the quadrupole operators $\{J^{\pm},J^z\}$ in this manifold. These have off-diagonal matrix elements, and are identified with the pseudo-spin raising and lowering operators ($\sigma^\pm=\sigma^1\pm i\sigma^2$). 

\begin{figure}
\centering
\includegraphics[scale=0.5]{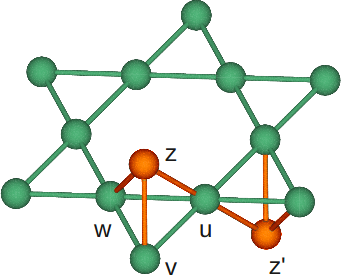}
\caption{A Kagome layer, in the pyrochlore lattice environment. We consider sites labelled z and z' replaced by non-magnetic ions, decoupling the Kagome layers. The local axis at the u,v and w sites point towards the center of the tetrahedron on which these lie.}
\label{fig_3dkagome}
\end{figure}

In a pyrochlore lattice the local $D_{3d}$ axes point to the centre of the tetrahedra.\cite{2011_onada} On looking at the pyrochlore lattice along the [111] direction, it is found to be made out of alternate layers of Kagome and triangular lattices. For each Kagome layer (shown in Fig. \ref{fig_3dkagome}) the local $D_{3d}$ axes make an angle of $\cos^{-1}(\sqrt{2/3})$  with the plane of the Kagome layer. We imagine replacing the Pr$^{3+}$ ions from the triangular lattice layer with non-magnetic ions so as to obtain decoupled Kagome layers with Pr$^{3+}$ ions on the sites. The resulting structure is obtained in the same spirit as the now well-known Kagome compound Herbertsmithite was envisioned. As long as the local crystal field has $D_{3d}$ symmetry, the doublet remains well defined. A suitable candidate non-magnetic ion may be iso-valent but non-magnetic La$^{3+}$. Notice that the most extended orbitals in both cases are the fifth shell orbitals and the crystal field at each Pr$^{3+}$ site is mainly determined by the surrounding oxygens and the transition metal element. Hence, we expect that the splitting of the non-Kramers doublet due to the above substitution would be very small and the doublet will remain well defined. In this work we shall consider such a Kagome lattice layer and analyze possible $Z_2$ spin-orbital liquids, with gapped or gapless fermionic {\it spinons}.

The rest of the paper is organized as follows. In Sec. \ref{sec_symm_ham}, we begin with a discussion of the symmetries of the non-Kramers system on a Kagome lattice and write down the most general pseudo-spin model with pseudo-spin exchange interactions up to second nearest neighbours. In Sec. \ref{sec_spinon_psg} formulate the projective symmetry group (PSG) analysis for singlet and triplet decouplings. Using this we demonstrate that the non-Kramers transformation of our pseudo-spin degrees of freedom under time reversal leads to a set of ten spin-orbital liquids which cannot be realized in the Kramers case. In Sec. \ref{sec_struct} we derive the dynamic spin-spin structure factor for a representative spin liquid for the case of both Kramers and non-Kramers doublets, demonstrating that experimentally measurable properties of these two types of spin-orbital liquids differ qualitatively. Finally, in Sec. \ref{sec_discuss}, we discuss our results, and propose an experimental test which can detect a non-Kramers spin-orbital liquid. The details of various calculations are discussed in different appendices.

\section{Symmetries and the pseudo-spin Hamiltonian}
\label{sec_symm_ham}

\begin{figure}
\centering
\includegraphics[scale=0.75]{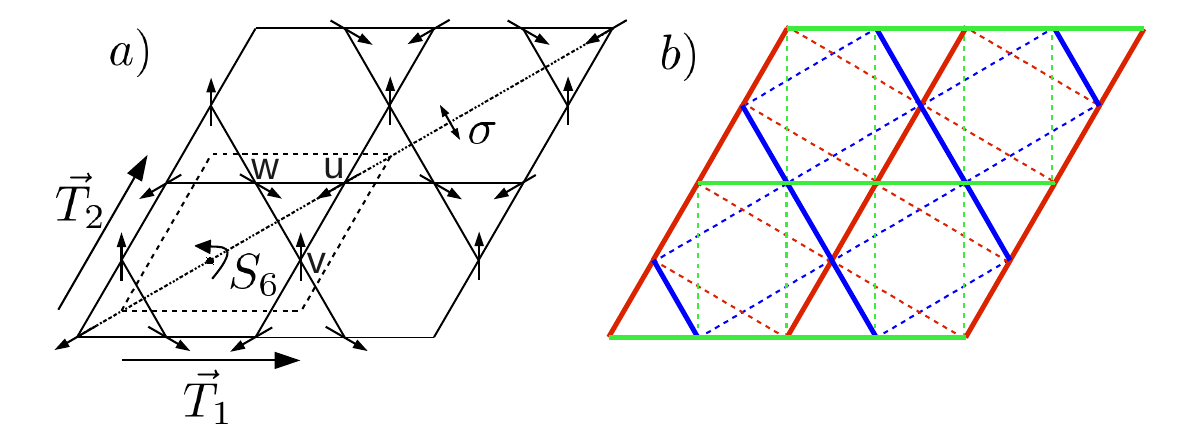}
\caption{(color online) (a) The symmetries of the Kagome lattice. Also shown are the labels for the sublattices and the orientation of the local z-axis. (b) Nearest and next nearest neighbour bonds. Colors refer to the phases $\phi_{r,r'}$ and $\phi '_{r,r'}$, with these being 0 on blue bonds, 1 on green bonds and 2 on red bonds.}
\label{fig_symms}
\end{figure}

Since the local $D_{3d}$ axes of the three sites in the Kagome unit cell differ from each other a general pseudo-spin Hamiltonian is not symmetric under {\it continuous} global pseudo-spin rotations. However, it is symmetric under various symmetry transformations of the Kagome lattice as well as time reversal symmetry. Such symmetry transformations play a major role in the remainder of our analysis. We start by describing the effect of various lattice symmetry transformations on the non-Kramers doublet.

We consider the symmetry operations that generate the space group of the above Kagome lattice. These are (as shown in Fig. \ref{fig_symms}(a))
\begin{itemize}
\item $T_1$, $T_2$ : generate the two lattice translations.
\item $\quad \sigma=C'_2 I$ : (not to be confused with the pseudo-spin operators which come with a superscript)  where $I$ is the three dimensional inversion operator about a plaquette center and   $C'_2$ refers to a two-fold rotation about a line joining two opposite sites on the plaquette.
\item $\quad S_6=C_3^2 I$ : where  $C_3$ is the threefold rotation operator about the center of a hexagonal plaquette of the Kagome lattice.
\item $\quad T=\sigma_1 K$ :  Time reversal.
\end{itemize}
Here, we consider a three dimensional inversion operator since the local $D_{3d}$ axes point out of the Kagome plane. The above symmetries act non-trivially on the pseudo-spin degrees of freedom, as well as the lattice degrees of freedom. The action of the symmetry transformations on the pseudo-spin operators is given by,
\begin{align}
S_6: \{\sigma^3,\sigma^+,\sigma^-\} &\rightarrow \{\sigma^3,\bar{\omega} \sigma^+, \omega\sigma^-\},\nonumber\\
T: \{\sigma^3,\sigma^+,\sigma^-\} &\rightarrow \{-\sigma^3,\sigma^-,\sigma^+\},\nonumber\\
{C'_2} : \{\sigma^3,\sigma^+,\sigma^-\} &\rightarrow \{-\sigma^3,\sigma^-,\sigma^+\},\nonumber\\
T_1: \{\sigma^3,\sigma^+,\sigma^-\} &\rightarrow \{\sigma^3,\sigma^+,\sigma^-\}, \nonumber\\
T_2: \{\sigma^3,\sigma^+,\sigma^-\} &\rightarrow \{\sigma^3,\sigma^+,\sigma^-\},
\label{eq_orbtrans}
\end{align}
($\omega=\bar{\omega}^{-1}=e^{i\frac{2\pi}{3}}$). Operationally their action on the doublet $(|+\rangle ~~|-\rangle)$ can be written in form of $2\times 2$ matrices. The translations $T_1,$ $T_2$ act trivially on the pseudo-spin degrees of freedom, and the remaining operators act as
\begin{align}
T = \sigma_1 K, \qquad \sigma = \sigma_1,~~~~
S_6 = \begin{bmatrix}
\bar{\omega} & 0\\
0 & \omega
\end{bmatrix},
\label{eq_pseudoops}
\end{align}
where $K$ refers to complex conjugation. The above expressions can be derived by examining the effect of these operators on the wave-function describing the doublet (Eq. \ref{eq_wavefn}).

We can now write down the most generic pseudo-spin Hamiltonian allowed by the above lattice symmetries that is bilinear in pseudo-spin operators. The form of the time-reversal symmetry restricts our attention to those products which are formed by a pair of $\sigma^3$ operators or those which mix the pseudo-spin raising and lowering operators. Any term which mixes $\sigma^3$ and $\sigma^\pm$ changes sign under the symmetry, and
 can thus be excluded. Under the $C_3$ transformation about a site, the terms $C_3: \sigma^3_r \sigma^3_{r'}\rightarrow \sigma^3_{C_3(r)} \sigma^3_{C_3(r')}$ and $C_3:\sigma^+_r \sigma^-_{r'}\rightarrow \sigma^+_{C_3(r)} \sigma^-_{C_3(r')}$. However, the term $\sigma^+_r \sigma^+_{r'}$ (and its Hermitian conjugate) gain additional phase factors when transformed; under the $C_3$ symmetry transformation, this term becomes $C_3:\sigma^+_{r} \sigma^+_{r'}\rightarrow\bar{\omega}\sigma^+_{C_3(r)} \sigma^+_{C_3(r')}$. In addition, under the $\sigma$ symmetry, this term transforms as $\sigma : \sigma^+_r \sigma^+_{r'} \rightarrow \sigma^-_{\sigma(r)}\sigma^-_{\sigma(r')}$. Thus the Hamiltonian with spin-spin exchange interactions up to next-nearest neighbour is given by
\begin{align}
H_{eff} = J_{nn}\sum_{\langle r,r'\rangle} [&\sigma_{r}^3\sigma_{r'}^3 + 2(\delta \sigma_r^+\sigma_{r'}^- + h.c.)\nonumber\\
+& 2q(e^{\frac{2 \pi i\phi_{r,r'}}{3}}\sigma_r^+\sigma_{r'}^+ + h.c.)]\nonumber\\
+& J_{nnn}\sum_{\langle \langle r,r' \rangle \rangle} [\sigma_{r}^3\sigma_{r'}^3 + 2(\delta ' \sigma_r^+\sigma_{r'}^- + h.c.)\nonumber\\
+& 2q'(e^{\frac{2 \pi i\phi '_{r,r'}}{3}}\sigma_r^+\sigma_{r'}^+ + h.c.)],
\label{eq_spinham}
\end{align}
where $\phi$ and $\phi '$ take values 0, 1 and 2 depending on the bonds on which they are defined (Fig. \ref{fig_symms}(b)).

\section{Spinon  representation of the pseudo-spins and PSG analysis}
\label{sec_spinon_psg}

Having written down the pseudo-spin Hamiltonian, we now discuss the possible spin-orbital liquid phases. We do this in two stages in the following sub-sections.
\subsection{Slave fermion representation and spinon decoupling}

In order to understand these phases, we will use the fermionic slave-particle decomposition of the pseudo-spin operators. At this point, we note that the pseudo-spins satisfy $S=1/2$ representations of a ``SU(2)" algebra among their generators (not to be confused with the regular spin rotation symmetry). We represent the pseudo-spin degrees of freedom in terms of a fermion bilinear. This is very similar to usual slave fermion construction for spin liquids\cite{2002_wen,1987_anderson}. We take
\beq
\sigma_{j}^\mu = \frac{1}{2} f_{j\alpha}^\dag [\rho^\mu]_{\alpha \beta}f_{j\beta},
\eeq
where $\alpha,\beta=\uparrow,\downarrow$ is defined along the local $z$ axis and $f^\dag$ ($f$) is an $S=1/2$ fermionic creation (annihilation) operator. Following standard nomenclature, we refer to the $f (f^\dagger)$ as the spinon annihilation (creation) operator, and note that these satisfy standard fermionic anti-commutation relations. The above spinon representation, along with the single occupancy constraint
\beq
 f^\dagger_{i\uparrow}f_{i\uparrow}+f^\dagger_{i\downarrow}f_{i\downarrow}=1,
 \label{eq_singoccupancydirac}
\eeq
form a faithful representation of the pseudo-spin-1/2 Hilbert space. The above representation of the pseudo-spins, when used in Eq. \ref{eq_spinham}, leads to a quartic spinon Hamiltonian. Following standard procedure,\cite{2002_wen,1987_anderson} this is then decomposed using auxiliary fields into a quadratic spinon Hamiltonian (after writing down the corresponding Eucledian action). The mean field description of the phases is then characterized by the possible saddle point values of the auxiliary fields. There are eight such auxiliary fields per bond, corresponding to
\begin{subequations}
\begin{align}
\chi_{ij}=\langle f^\dagger_{i\alpha}f_{j\alpha}\rangle^*;~~~~~\eta_{ij}=\langle f_{i\alpha}\left[i\tau^2\right]_{\alpha\beta}f_{j\beta}\rangle^*;
\label{eq_singans}
\end{align}
\begin{align}
E^{a}_{ij}=\langle f^\dagger_{i\alpha}\left[\tau^a\right]_{\alpha\beta}f_{j\beta}\rangle^*;~~~~D^a_{ij}=\langle f_{i\alpha}\left[i\tau^2\tau^a\right]_{\alpha\beta}f_{i\beta}\rangle^*;
\label{eq_tripans}
\end{align}
\end{subequations}
where $\tau^a$ ($a=1,2,3$) are the Pauli matrices. While Eq. \ref{eq_singans} represents the usual singlet spinon hopping (particle-hole) and pairing (particle-particle) channels, Eq. \ref{eq_tripans} represents the corresponding triplet decoupling channels. Since the Hamiltonian (Eq. \ref{eq_spinham}) does not have pseudo-spin rotation symmetry, both the singlet and the triplet decouplings are necessary.\cite{2013_dodds,2012_schaffer}

From this decoupling, we obtain a mean-field Hamiltonian which is quadratic in the spinon operators. We write this compactly in the following form\cite{2012_schaffer} (subject to the constraint Eq.\ref{eq_singoccupancydirac})
\begin{align}
H_0 &= \sum_{ij}J_{ij}\vec{f}_i^\dag U_{ij} \vec{f}_j, \\
\vec{f}_i^\dag &= \begin{bmatrix}
f_{i\uparrow}^\dag & f_{i\downarrow} & f_{i\downarrow}^\dag & -f_{i\uparrow}
\end{bmatrix},\\
U_{ij} &= \xi_{ij}^{\alpha\beta}\Sigma^\alpha \Gamma^\beta, \\
\Sigma^\alpha &= \rho^\alpha \otimes I; \qquad \Gamma^\beta = I \otimes \tau^\beta,
\label{eq_hmf}
\end{align}
where $\rho^\alpha$ are the Identity (for $\alpha=0$) and Pauli matrices ($\alpha=1,2,3$) acting on pseudo-spin degrees of freedom, and $\tau^\alpha$ represents the same in the gauge space. We immediately note that
\begin{align}
\left[\Sigma^\alpha,\Gamma^\beta\right]=0~~~~~\forall\alpha,\beta .
\label{eq_commute}
\end{align}
The requirement that our $H_0$ be Hermitian restricts the coefficients $\xi_{ij}$ to satisfy
\begin{align}
\xi_{ij}^{00},\xi_{ij}^{ab} \in \Im;~~~~\xi_{ij}^{a0},\xi_{ij}^{0b} \in \Re.
\end{align}
for $a,b \in \{1,2,3\}$. The relations between $\xi_{ij}$s and $\{\chi_{ij},\eta_{ij},{\bf E}_{ij}, {\bf D}_{ij}\}$ are given in Appendix \ref{appen_param_relation}.\cite{2012_schaffer}
As a straight forward extension of the $SU(2)$ gauge theory formulation for spin liquids,\cite{1988_affleck,2002_wen} we find that $H_0$ is invariant under the gauge transformation
\begin{align}
\vec{f}_j \rightarrow W_j \vec{f}_j,\\
U_{ij} \rightarrow W_i U_{ij} W_j^\dag ,
\label{eq_gauge_trans}
\end{align}
where the $W_i$ matrices are SU(2) matrices of the form $W_i = e^{i\vec{\Gamma}\cdot \vec{a_i}}$ $(\vec\Gamma\equiv(\Gamma^1,\Gamma^2,\Gamma^3))$. Noting that the physical pseudo-spin operators are given by
\begin{align}
{\vec \sigma}_i = \frac{1}{4}\vec{f}_i^\dag \vec{\Sigma} \vec{f}_i,
\end{align}
Eq. \ref{eq_commute} shows that the spin operators, as expected, are gauge invariant. It is useful to define the ``$\Sigma$-components" of the $U_{ij}$ matrices as follows:
\begin{align}
U_{ij}=\mathcal{V}_{ij}^\alpha\Sigma^\alpha ,
\end{align}
where
\begin{align}
\mathcal{V}_{ij}^\alpha=\xi_{ij}^{\alpha \beta}\Gamma^\beta=\left[\begin{array}{cc}
\mathcal{J}_{ij}^\alpha & 0\\
0 & \mathcal{J}_{ij}^\alpha\end{array}\right],
\end{align}
and
\begin{align}
\mathcal{J}_{ij}^\alpha=\left[\begin{array}{cc}
\xi_{ij}^{\alpha0}+\xi_{ij}^{\alpha3} & \xi_{ij}^{\alpha1}-i \xi_{ij}^{\alpha2}\\
 \xi_{ij}^{\alpha1}+i \xi_{ij}^{\alpha2} & \xi_{ij}^{\alpha0}-\xi_{ij}^{\alpha3}
\end{array}\right].
\end{align}
Under global spin rotations the fermions transform as
\beq
\vec{f}_i \rightarrow V \vec{f}_i ,
\eeq
where V is an SU(2) matrix of the form $V = e^{i\vec{\Sigma}\cdot \vec{b}}$ ($\vec{\Sigma}\equiv\{\Sigma^1,\Sigma^2,\Sigma^3\}$). So while $\mathcal{V}^0_{ij}$ (the singlet hopping and pairing) is invariant under spin rotation, $\{\mathcal{V}^1_{ij}, \mathcal{V}^2_{ij},\mathcal{V}^3_{ij}\}$ transforms as a vector as expected since they represent triplet hopping and pairing amplitudes.

\subsection{PSG Classification}

We now classify the non-Kramers spin-orbital liquids based on projective representation similar to that of the conventional quantum spin liquids.\cite{2002_wen} Each spin-orbital liquid ground state of the quadratic Hamiltonian (Eq \ref{eq_hmf}) is characterized by the mean field parameters (eight on each bond, $\chi,\eta,E^1,E^2,E^3,D^2,D^2,D^3$, or equivalently $U_{ij}$). However, due to the gauge redundancy of the spinon parametrization (as shown in Eq. \ref{eq_gauge_trans}), a general mean-field ansatz need not be invariant under the symmetry transformations on their own but may be transformed to a gauge equivalent form without breaking the symmetry. Therefore, we must consider its transformation properties under a projective representation of the symmetry group.\cite{2002_wen} For this, we need to know the various projective representations of the lattice symmetries  of the Hamiltonian (Eq. \ref{eq_spinham}) in order to classify different spin-orbital liquid states.

Operationally, we need to find different possible sets of gauge transformations $\{GG\}$ which act in combination with the symmetry transformations $\{SG\}$ such that the mean-field ansatz $U_{ij}$ is invariant under such a combined transformation. In the case of spin rotation invariant spin-liquids (where only the singlet channels $\chi$ and $\eta$ are present), the above statement is equivalent to demanding the following invariance:
\begin{align}
U_{ij}=\left[G_S S\right]U_{ij}\left[G_S S\right]^\dagger=G_S(i)U_{S(i)S(j)}G_S^\dagger(j) ,
\label{eq_uijusual}
\end{align}
where $S \in SG$ is a symmetry transformation and $G_S\in GG$ is the corresponding gauge transformation. The different possible $\{G_S |\forall S\in SG\}$ give the possible algebraic PSGs that can characterize the different spin-orbital liquid phases. To obtain the different PSGs, we start with various lattice symmetries of the Hamiltonian. The action of various lattice transformations\cite{2011_lu} is given by
\begin{align}
T_1:& (x,y,s) \rightarrow (x+1,y,s);\nonumber\\
T_2:& (x,y,s) \rightarrow (x,y+1,s);\nonumber\\
\sigma:& (x,y,u) \rightarrow (y,x,u);\nonumber\\
&(x,y,v) \rightarrow (y,x,w);\nonumber\\
&(x,y,w) \rightarrow (y,x,v);\nonumber\\
S_6:& (x,y,u) \rightarrow (-y-1,x+y+1,v);\nonumber\\
&(x,y,v) \rightarrow (-y,x+y,w);\nonumber\\
&(x,y,w) \rightarrow (-y-1,x+y,u);
\end{align}
where $(x,y)$ denotes the lattice coordinates and $s\in \{u,v,w\}$ denotes the sub-lattice index (see figure \ref{fig_symms}).

In terms of the symmetries of the Kagome lattice, these operators obey the following conditions
\begin{gather}
T^2=\sigma^2 = (S_6)^6 = e,\nonumber\\
g^{-1}T^{-1}gT = e \quad \forall g \in SG,\nonumber\\
T_2^{-1}T_1^{-1}T_2T_1 = e,\nonumber\\
\sigma^{-1}T_1^{-1}\sigma T_2 = e,\nonumber\\
\sigma^{-1}T_2^{-1}\sigma T_1 = e,\nonumber\\
S_6^{-1}T_2^{-1}S_6T_1 = e,\nonumber\\
S_6^{-1}T_2^{-1}T_1S_6T_2 = e,\nonumber\\
\sigma^{-1}S_6\sigma S_6 = e.
\label{eq_commrels}
\end{gather}
In addition, these commutation relations are valid in terms of the operations on the pseudo-spin degrees of freedom, as can be verified from Eq. \ref{eq_pseudoops}.

In addition to the conditions in Eq. \ref{eq_commrels}, the Hamiltonian is trivially invariant under the identity transformation. The invariant gauge group (IGG) of an ansatz is defined as the set of all pure gauge transformations $G_I$ such that $G_I:U_{ij} \rightarrow U_{ij}$. The nature of such pure gauge transformations immediately dictates the nature of the low energy fluctuations about the mean field state. If these fluctuations do not destabilize the mean-field state, we get stable spin liquid phases whose low energy properties are controlled by the IGG. Accordingly, spin liquids obtained within projective classification are primarily labelled by their IGGs and we have $Z_2, U(1)$ and $SU(2)$ spin liquids corresponding to IGGs of $Z_2, U(1)$ and $SU(2)$ respectively. In this work we concentrate on the set of $Z_2$ ``spin liquids" (spin-orbital liquids with a $Z_2$ IGG).

We now focus on the PSG classification. As shown in Eq. \ref{eq_orbtrans}, in the present case, the pseudo-spins transform non-trivially under different lattice symmetry transformations. Due to the presence of the triplet decoupling channels the non-Kramers doublet transforms non-trivially under lattice symmetries (Eq. \ref{eq_pseudoops}). Thus, the invariance condition on the $U_{ij}$s is not given by Eq. \ref{eq_uijusual}, but by a more general condition
\beq
U_{ij}=\left[G_S S\right] U_{ij} \left[G_S S\right]^\dagger= G_S(i)\phi_S\left[U_{S(i)S(j)}\right]G_S^\dagger(j).
\eeq
Here
\begin{align}
\phi_S\left[U_{S(i)S(j)}\right]=\mathcal{D}_S U_{S(i)S(j)}\mathcal{D}_S^\dagger ,
\label{eq_spintrans2}
\end{align}
and $\mathcal{D}_S$ generates the pseudo-spin rotation associated with the symmetry transformation ($S$) on the doublet. The matrices $\mathcal{D}_S$ have the form
\begin{gather}
\mathcal{D}_{S_6} = -\frac{1}{2}\Sigma^0 -\frac{i \sqrt{3}}{2} \Sigma^3 ,\\
\mathcal{D}_{\sigma} = \mathcal{D}_{T} =  i*\Sigma^1 , \qquad \mathcal{D}_{T_1} = \mathcal{D}_{T_2} = \Sigma^0 .
\end{gather}

Under these constraints, we must determine the relations between the gauge transformation matrices $G_S(i)$ for our set of ansatz. The additional spin transformation (Eq. \ref{eq_spintrans2}) does not affect the structure of the gauge transformations, as the gauge and spin portions of our ansatz are naturally separate (Eq. \ref{eq_commute}). In particular, we can choose to define our gauge transformations such that
\begin{align}
G_S: U_{ij} &= G_S: \xi_{ij}^{\alpha\beta}\Sigma^\alpha \Gamma^\beta \rightarrow \xi_{ij}^{\alpha\beta}\Sigma^\alpha G_S^\dag(i) \Gamma^\beta G_S(j) ,\\
S: U_{ij} &= S: \xi_{ij}^{\alpha\beta}\Sigma^\alpha \Gamma^\beta \rightarrow \xi_{S(i)S(j)}^{\alpha\beta} \mathcal{D}_S \Sigma^\alpha \mathcal{D}_S^\dag \Gamma^\beta,
\end{align}
where we have used the notation $G_S: U_{ij}\equiv G_S^\dagger(i)U_{ij}G_S(j)$ and so forth. As a result, we can build on the general construction of Lu {\it et al.}\cite{2011_lu} to derive the form of the gauge transformation matrices. The details are given in Appendix \ref{app_gauge}.

A major difference arises when examining the set of algebraic PSGs for $Z_2$ spin liquids found on the Kagome lattice due to the difference between the structure of the time reversal symmetry operation on the Kramers and non-Kramers pseudo-spin-$1/2$s. In the present case, we find there are 30 invariant PSGs leading to {\it thirty} possible spin-orbital liquids. This is in contrast with the Kramers case analysed by Lu {\it et al.},\cite{2011_lu} where  {\it ten} of the algebraic PSGs cannot be realized as invariant PSGs, as all bonds in these ansatz are predicted to vanish identically due to the form of the time reversal operator, and hence there are only twenty possible spin liquids. However, with the inclusion of spin triplet terms and the non-Kramers form of our time reversal operator, these ansatz are now realizable as invariant PSGs as well. The time reversal operator, as defined in Appendix \ref{app_gauge}, acts as
\beq
T: \xi_{ij}^{\alpha \beta} \Sigma^\alpha \Gamma^\beta \rightarrow \tilde{\xi}_{ij}^{\alpha \beta} \Sigma^\alpha \Gamma^\beta,
\eeq
where $\tilde{\xi}^{\alpha \beta}$ = $\xi^{\alpha \beta}$ if $\alpha \in \{1,2\}$ and $\tilde{\xi}^{\alpha \beta}$ = $-\xi^{\alpha \beta}$ if $\alpha \in \{0,3\}$. The projective implementation of the time-reversal symmetry condition (Eq. \ref{eq_commrels}) takes the form (see Appendix \ref{app_gauge})
\begin{align}
\left[G_T(i)\right]^2=\eta_T I~~~~~~\forall i ,
\end{align}
where $G_T(i)$ is the gauge transformation associated with time reversal operation and $\eta_T=\pm 1$ for a $Z_2$ IGG.

Therefore, the terms allowed by the time reversal symmetry to be non zero are, for $\eta_{T} = 1$,
\begin{align}
\label{eq_eta1}
\xi^{10},\xi^{11},\xi^{12},\xi^{13},\xi^{20},\xi^{21},\xi^{22},\xi^{23},
\end{align}
and for $\eta_{T} = -1$, with the choice $G_T(i) = i\Gamma^1$ (see appendix \ref{app_gauge}),
\begin{align}
\label{eq_eta-1}
\xi^{02},\xi^{03},\xi^{10},\xi^{11},\xi^{20},\xi^{21},\xi^{32},\xi^{33}.
\end{align}
This contrasts with the case of Kramers doublets, in which no terms are allowed for $\eta_{T} = -1$, and for $\eta_{T}=-1$ the allowed terms are
\begin{align}
\xi^{02},\xi^{03},\xi^{12},\xi^{13},\xi^{22},\xi^{23},\xi^{32},\xi^{33}.
\end{align}

\begin{table*}
\caption{Symmetry allowed terms: We list the terms allowed to be non-zero by symmetry, for the 30 PSGs determined by Yuan-Ming Lu {\it et al}\cite{2011_lu}. The PSGs listed together are those with $\eta_{12} = \pm 1$ and all other factors equal. Included are terms allowed on nearest and next-nearest neighbour bonds, as well as chemical potential terms $\Gamma$ which can be non zero on all sites for certain spin-orbital liquids. Also included is the distance of bond up to which we must include in order to gap out the gauge fluctuations to $Z_2$ via the Anderson-Higgs mechanism\cite{2002_wen}. Only PSGs 9 and 10 can not host $Z_2$ spin-orbital liquids with up to second nearest neighbour bonds.}
\begin{center}
{\renewcommand{\arraystretch}{1.5}
\renewcommand{\tabcolsep}{0.2cm}
	\begin{tabular}{c|c|c|c|c}
		\hline
		No. & $\Lambda_s$ & n.n. & n.n.n. & $Z_2$ \\ \hline
		1-2 & $\Gamma^2,\Gamma^3$ & $\xi^{10},\xi^{21},\xi^{02},\xi^{03},\xi^{32},\xi^{33}$ &$ \xi^{10},\xi^{21},\xi^{02},\xi^{03},\xi^{32},\xi^{33}$ & n.n. \\ \hline
		3-4 & 0 & $\xi^{10},\xi^{21},\xi^{02},\xi^{03},\xi^{32},\xi^{33} $&$ \xi^{10},\xi^{21} $& n.n. \\ \hline
		5-6 & $\Gamma^3$&$ \xi^{10},\xi^{21},\xi^{02},\xi^{03},\xi^{32},\xi^{33} $& $\xi^{10},\xi^{21},\xi^{03},\xi^{33} $& n.n. \\ \hline
		7-8 & 0 &$ \xi^{11},\xi^{20} $& $\xi^{11},\xi^{20},\xi^{02},\xi^{03},\xi^{32},\xi^{33}$ & n.n.n. \\ \hline
		9-10 & 0 &$ \xi^{11},\xi^{20} $&$ \xi^{11},\xi^{20} $& - \\ \hline
		11-12 & 0 &$ \xi^{11},\xi^{20} $&$ \xi^{10},\xi^{11},\xi^{02},\xi^{32}$ & n.n.n. \\ \hline
		13-14 &$ \Gamma^3 $&$ \xi^{10},\xi^{11},\xi^{03},\xi^{33} $&$ \xi^{10},\xi^{21},\xi^{02},\xi^{03},\xi^{32},\xi^{33}$ & n.n. \\ \hline
		15-16 & $\Gamma^3 $& $\xi^{10},\xi^{11},\xi^{03},\xi^{33}$ & $\xi^{10},\xi^{21},\xi^{03},\xi^{33} $& n.n. \\ \hline
		17-18 & 0 &$ \xi^{10},\xi^{11},\xi^{03},\xi^{33}$ & $\xi^{10},\xi^{11},\xi^{02},\xi^{32}$ & n.n. \\ \hline
		19-20 & 0 &$ \xi^{10},\xi^{11},\xi^{03},\xi^{33} $&$ \xi^{10},\xi^{21} $& n.n. \\ \hline
		21-22 & 0 &$ \xi^{10},\xi^{21},\xi^{22},\xi^{23}$ &$ \xi^{10},\xi^{21},\xi^{22},\xi^{23} $& n.n.n. \\ \hline
		23-24 & 0 &$ \xi^{10},\xi^{21},\xi^{22},\xi^{23} $&$ \xi^{10},\xi^{11},\xi^{12},\xi^{23}$ & n.n. \\ \hline
		25-26 & 0 &$ \xi^{11},\xi^{12},\xi^{13},\xi^{20} $&$ \xi^{13},\xi^{20},\xi^{21},\xi^{22} $& n.n. \\ \hline
		27-28 & 0 &$ \xi^{11},\xi^{12},\xi^{13},\xi^{20}$ &$ \xi^{10},\xi^{11},\xi^{13},\xi^{22} $& n.n. \\ \hline
		29-30 & 0 & $\xi^{11},\xi^{12},\xi^{13},\xi^{20}$ & $\xi^{11},\xi^{12},\xi^{13},\xi^{20} $& n.n. \\ \hline
	\end{tabular}}
\end{center}
\label{tab_allowed}
\end{table*}

Further restrictions on the allowed terms on each link arise from the form of the gauge transformations defined for the symmetry transformations. All nearest neighbour bonds can then be generated from $U_{ij}$ defined on a single bond, by performing appropriate symmetry operations.

Using the methods outlined in earlier works (Ref. \onlinecite{2002_wen}, \onlinecite{2011_lu}) we find the {\it minimum} set of parameters required to stabilize $Z_2$ spin-orbital liquids. We take into consideration up to second neighbour hopping and pairing amplitudes (both singlet and triplet channels). The results are listed in Table \ref{tab_allowed}.

The spin-orbital liquids listed from $21-30$ are not allowed in the case of Kramers doublets and, as pointed out before, their existence is solely due to the unusual action of the time-reversal symmetry operator on the non-Kramers spins. Hence these {\it ten} spin-orbital liquids are qualitatively new phases that may appear in these systems. Of these ten phases, only {\it two} (labelled as $21$ and $22$ in Table \ref{tab_allowed}) require next nearest neighbour amplitudes to obtain a $Z_2$ spin-orbital liquid. For the other {\it eight}, nearest neighbour amplitudes are already sufficient to stabilize a $Z_2$ spin-orbital liquid.

It is interesting to note (see below) that {\it bond-pseudo-spin-nematic} order (Eq. \ref{eq_scnem} and Eq. \ref{eq_venem}) can signal spontaneous time-reversal symmetry breaking. Generally, since the triplet decouplings are present,  the bond nematic order parameter for the pseudo-spins\cite{2009_shindou,2012_bhattacharjee}
\begin{align}
\mathcal{Q}^{\alpha\beta}_{ij}=\langle\left(S_i^\alpha S_j^\beta+S_i^\beta S_j^\alpha\right)/2-\delta^{\alpha\beta}(\vec S_i\cdot\vec S_j)/3\rangle,
\label{eq_scnem}
\end{align}
as well as vector chirality order
\begin{align}
\vec{\mathcal{J}}_{ij}=\langle\vec S_i\times \vec S_j\rangle,
\label{eq_venem}
\end{align}
are non zero.  Since the underlying Hamiltonian Eq. \ref{eq_spinham}) generally does not have pseudo-spin rotation symmetry, the above non-zero expectation values do not spontaneously break any pseudo-spin rotation symmetry. However, because of the unusual transformation property of the non-Kramers pseudo-spins under time reversal, the operators corresponding to $\mathcal{Q}^{13}_{ij},\mathcal{Q}^{23}_{ij},\mathcal{J}^1_{ij},\mathcal{J}^2_{ij}$ are odd under time reversal, a symmetry of the pseudo-spin Hamiltonian. Hence if any of the above operators gain a non-zero expectation value in the ground state, then the corresponding spin-orbital liquid breaks time reversal symmetry. While this can occur in principle, we check explicitly (see Appendix \ref{appen_param_relation}) that in all the spin-orbital liquids discussed above, the expectation values of these operators are identically zero. This provides a non-trivial consistency check on our PSG calculations.

We now briefly dicuss the effect of the fluctuations about the mean-field states. In the absence of pairing channels (both singlet and triplet) the  gauge group is $U(1)$. In this case, the fluctuations of the gauge field about the mean field (Eq. \ref{eq_gauge_trans}) are related to the scalar pseudo-spin chirality $\vec S_1\cdot \vec S_2\times \vec S_3$, where the three sites form a triangle.\cite{1992_lee} Such fluctuations are gapless in a $U(1)$ spin liquid. It is interesting to note that the scalar spin-chirality is odd under time-reversal symmetry and it has been proposed that such fluctuations can be detected in neutron scattering experiments in presence of spin rotation symmetry breaking.\cite{2013_lee} In the present case, however, due to the presence of spinon pairing, the gauge group is broken down to $Z_2$ and the above gauge fluctuations are rendered gapped through Anderson-Higg's mechanism.\cite{2002_wen}

In addition to the above gauge fluctuations, because of the triplet decouplings which break pseudo-spin rotational symmetry, there are bond quadrupolar fluctuations of the pseudo-spins $\mathcal{Q}^{\alpha\beta}_{ij}$ (Eq. \ref{eq_scnem}), as well as vector chirality fluctuations $\vec{\mathcal{J}}_{ij}$ (Eq. \ref{eq_venem})\cite{2009_shindou,2012_bhattacharjee} on the bonds. These nematic and vector chirality fluctuations are gapped because the underlying pseudo-spin Hamiltonian (Eq. \ref{eq_spinham}) breaks pseudo-spin-rotation symmetry. However, we note that because of the unusual transformation of the non-Kramers pseudo-spins under time reversal (only the $z-$component of pseudo-spins being odd under time reversal), $\mathcal{Q}^{13}_{ij}, \mathcal{Q}^{23}_{ij}, \mathcal{J}^1_{ij}$ and $\mathcal{J}^2_{ij}$ are odd under time reversal. Hence, while their mean field expectation values are zero (see above), the fluctuations of these quantities can in principle linearly couple to the neutrons in addition to the $z-$component of the pseudo-spins.

Having identified the possible $Z_2$ spin-orbital liquids, we can now study typical dynamic structure factors for these spin-orbital liquids. In the next section we examine the typical spinon band structure for different spin-orbital liquids obtained above and find their dynamic spin structure factor.

\section{Dynamic spin structure factor}
\label{sec_struct}

We compute the dynamic spin structure factor
\begin{align}
S(q,\omega) = \int \frac{dt}{2\pi} e^{i\omega t} \sum_{ij}e^{i\bm{q}\cdot (\bm{r_i} - \bm{r_j})}\sum_{a=1,2,3}\langle \sigma_{i}^a(t) \sigma_j^a(0) \rangle,
\label{eq_struct}
\end{align}
for an example ansatz of our spin liquid candidates, in order to demonstrate the qualitative differences between the Kramers and non-Kramers spin-orbital liquids. In the above equation, the pseudo-spin variables are defined in a global basis (with the z-axis perpendicular to the Kagome plane). In computing the structure factor for the non-Kramers example, we include only the $\sigma^3$ components of the pseudo-spin operator in the local basis, since only the $z$-components carry magnetic dipole moment (see discussion before). Hence, only this component couples linearly to neutrons in a neutron scattering experiment.

\begin{figure}
\centering
\includegraphics[scale=0.45]{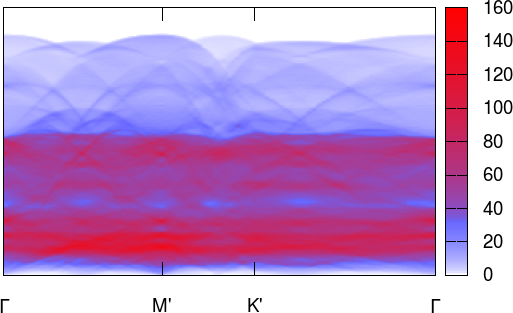}
\caption{The spin structure factor for an ansatz in spin liquid 17, with the spin variables transforming as a Kramers doublet.}
\label{fig_rotate}
\end{figure}
\begin{figure}
\centering
\includegraphics[scale=0.45]{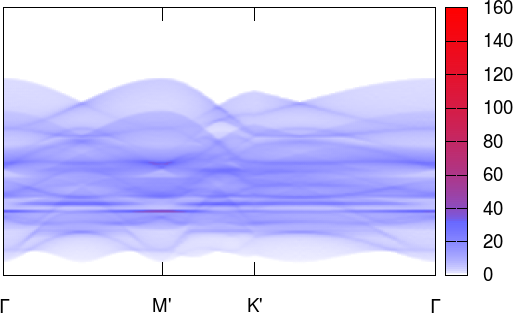}
\caption{The spin structure factor for an ansatz in spin liquid 17, with the spin variables transforming as a non-Kramers doublet.}
\label{fig_rotate}
\end{figure}

Eq. \ref{eq_struct} fails to be periodic in the first Brillouin zone of the Kagome lattice\cite{2013_dodds}, as the term $r_i - r_j$ in eq. \ref{eq_struct} is a half-integer multiple of the primitive lattice vectors when the sublattices of sites i and j are not equal. As such, we examine the structure factor in the extended brillouin zone, which consists of those momenta of length up to double that of those in the first brillouin zone. We plot the structure factor along the cut $\Gamma \rightarrow M' \rightarrow K' \rightarrow \Gamma$, where $M'$ = $2M$ and $K'$ = $2K$. We examine the structure factors for two ansatz of spin liquid \# 17 which has both Kramers and non-Kramers analogues.

As expected, we find that the structure factor has greater weight in the case of a Kramers spin liquid. This is partially due to the fact that the moment of the scattering particle couples with all components of the spin, rather than simply the $z$-component. In addition, we note that the presence of terms allowed in the non-Kramers spin-orbital liquid induce the formation of a gap, which is absent for the Kramers case with up to second nearest neighbour singlet and triplet terms in this particular spin-orbital liquid. Qualitative and quantitative differences such as these, which can be observed in these structure factors between Kramers and non-Kramers spin-orbital liquids, provides one possible distinguishing experimental signature of these states. We shall not pursue this in detail in the present work.

\section{Discussion and possible experimental signature of non-Kramers spin-orbital liquids}
\label{sec_discuss}

In this work, we have outlined the possible $Z_2$ spin-orbital liquids, with gapped or gapless fermionic spinons, that can be obtained in a system of non-Kramers pseudo-spin-1/2s on a Kagome lattice of Pr$^{+3}$ ions. We find a total of {\it thirty}, 10 more than in the case of corresponding Kramers system, allowed within PSG analysis in presence of time reversal symmetry. The larger number of spin-orbital liquids is a result of the difference in the action of the time-reversal operator, when realized projectively. We note that the spin-spin dynamic structure factor can bear important signatures of a non-Kramers spin-orbital liquid when compared to their Kramers counterparts. Our analysis of the number of invariant PSGs leading to possibly different spin-orbital liquids that may be realizable in other lattice geometries will form interesting future directions.

We now briefly discuss an experiment that can play an important role in determining non-Kramers spin-orbital liquids. Since the non-Kramers doublets are protected by crystalline symmetries, lattice strains can linearly couple to the pseudo-spins. As we discussed, the transverse ($x$ and $y$) components of the pseudo-spins $\{\sigma^1,\sigma^2\}$ carry quadrupolar moments and hence are even under the time reversal transformation. Further, they transform under an $E_g$ irreducible representation of the local $D_{3d}$ crystal field. Hence any lattice strain which has this symmetry can linearly couple to the above two transverse components. It turns out that in the crystal type that we are concerned, there is indeed such a mode related to the distortion of the oxygen octahedra. Symmetry considerations show that the linear coupling is of the form $E_{g1}\sigma^1+E_{g2}\sigma^2$ ($\{E_{g1},E_{g2}\}$ being the two components of the distortion in the local basis). The above mode is Raman active. For a spin-liquid, we expect that as the temperature is lowered, the spinons become more prominent as deconfined quasiparticles. So the Raman active phonon can efficiently decay into the spinons due to the above coupling channel. If the spin liquid is gapless, then this will lead to anomalous broadening of the above Raman mode as the temperature is lowered, which, if observed, can be an experimental signature of the non-Kramers spin-orbital liquid.  The above coupling is forbidden in Kramers doublets by time-reversal symmetry and hence no such anomalous broadening is expected.

\acknowledgements
We thank T. Dodds, SungBin Lee, A. Paramekanti and J. Rau for insightful discussions. This research was supported by the NSERC, CIFAR, and Centre for Quantum Materials at the University of Toronto.

\appendix

\section{Crystal Field Effects}
\label{app:D3d}
In this appendix, we explore the breaking of the $J=4$ spin degeneracy by the crystalline electric field. The oxygen and TM ions form a $D_{3d}$ local symmetry environment around the $Pr^{3+}$ ions, splitting the ground state degeneracy of the electrons. This symmetry group contains 6 classes of elements: $E$, $2C_3$, $3C'_2$, $i$, $2S_6$, and $3\sigma_d$, where the $C_3$ are rotations by $2\pi /3$ about the local z axis, the $C'_2$ are rotations by $\pi$ about axis perpendicular to the local z axis, $i$ is inversion, $S_6$ is a rotation by $4\pi /3$ combined with inversion and $\sigma_d$ is a reflection about the plane connecting one corner and the opposing plane, running through the $Pr$ molecule about which this is measured (or, equivalently, a rotation about the x axis combined with inversion). For our J=4 manifold, these have characters given by
\begin{align}
\chi^{(4)}(E) &= 2*4 + 1 = 9 = \chi^{(4)}(i)\\
\chi^{(4)}(C_3) &= \chi^{(4)}(\frac{2\pi}{3}) = \frac{\sin(3\pi)}{\sin(\pi /3)} = 0 = \chi^{(4)}(S_6)\\
\chi^{(4)}(\sigma_d) &= \chi^{(4)}(\pi) = \frac{\sin(9\pi /2)}{\sin(\pi /2)} = 1 = \chi^{(4)}(C'_2)
\end{align}
where the latter equalities are given by the fact that our J=4 manifold is inversion symmetric. Thus, decomposing this in terms of $D_{3d}$ irreps, our l=4 manifold splits into a sum of doublet and singlet manifolds as
\beq
\label{eq_reptheory}
\Gamma_{l=4} = 3E_g + 2A_{1g} + A_{2g}.
\eeq

To examine this further, we need to consider the matrix elements of the crystal field potential between the states of different angular momenta. We know that this potential must be invariant under all group operations of $D_{3d}$, so we can examine the transformation properties of individual matrix elements, $\langle m | V | m' \rangle$. Under the $C_3$ operation, these states of fixed m transform as
\beq
C_3 |m\rangle = e^{\frac{2\pi im}{3}}|m\rangle = \omega^m |m\rangle \qquad (\omega = e^{\frac{2\pi i}{3}})
\eeq
and thus the matrix elements transform as
\beq
C_3: \langle m | V | m' \rangle \rightarrow \langle m |(C_3)^{-1} V C_3| m' \rangle = \omega^{m'-m} \langle m | V | m' \rangle .
\eeq
By requiring that this matrix be invariant under this transformation, we can see that this potential only contains matrix elements for mixing of states which have the z-component of angular momentum which differ by 3. Thus, our eigenstates are mixtures of the $|m=4\rangle $, $|m=1\rangle $, and $|m=-2\rangle $ states, of the $|m=3\rangle $, $|m=0\rangle $, and $|m=-3\rangle $ states, and of the $|m=-4\rangle $, $|m=-1\rangle $, and $|m=2\rangle $ states.

In addition to this, we have the transformation properties
\beq
T|m\rangle = (-1)^m|-m\rangle
\eeq
and
\beq
\sigma |m\rangle = (-1)^m|-m\rangle
\eeq
(where the operators for time reversal and reflection are bolded for future clarity). Inversion acts trivially on these states, as we have total angular momentum even. Thus our time-reversal and lattice reflection (about one axis) symmetries give us doublet states of eigenstates $\alpha |m=4\rangle + \beta |m=1\rangle - \gamma|m=-2\rangle $ and $\alpha |m=-4\rangle - \beta |m=-1\rangle - \gamma|m=2\rangle $ (with $\alpha$, $\beta$, $\gamma \in \Re$ in order to respect the time reversal symmetry) for the three eigenstates of V in these sectors. The eigenstates of the $|m=3\rangle $, $|m=0\rangle $, and $|m=-3\rangle $ portion of V must therefore split into three singlet states, by our representation theory argument \ref{eq_reptheory}. Due to the expected strong Ising term in our potential, we expect the eigenstate with maximal J to be the ground state, meaning that to analyze the properties of this ground state we are interested in a single doublet state, one with large $\alpha$ (close to one). We will restrict ourselves to this manifold from this point forward, and define the two states in this doublet as
\begin{align}
|+\rangle &= \alpha |m=4\rangle + \beta |m=1\rangle - \gamma|m=-2\rangle \\
|-\rangle &= \alpha |m=-4\rangle - \beta |m=-1\rangle - \gamma|m=2\rangle .
\end{align}
We shall also refer to states of angular momentum $|m=n\rangle$ as $|n\rangle$ for simplicity of notation.

\section{Gauge transformations}
\label{app_gauge}

We begin by describing the action of time reversal on our ansatz. The operation is antiunitary, and must be combined with a spin transformation $\sigma^1$ in the case of non-Kramers doublets. As a result, the operation acts as $T: \xi_{ij}^{\alpha \beta} \Sigma^\alpha \Gamma^\beta \rightarrow \xi_{ij}^{\alpha \beta \ast} \Sigma^1 \Sigma^{\alpha \ast} \Sigma^1 \Gamma^{\beta \ast}$. However, we can simplify this considerably by performing a gauge transformation in addition to the above transformation, which yields the same transformation on any physical variables. The gauge transformation we perform is $i\Gamma^2$, which changes the form of the time reversal operation to $T: \xi_{ij}^{\alpha \beta} \Sigma^\alpha \Gamma^\beta \rightarrow \xi_{ij}^{\alpha \beta \ast} \Sigma^1 \Sigma^{\alpha \ast} \Sigma^1 \Gamma^2 \Gamma^{\beta \ast} \Gamma^2 = \tilde{\xi}_{ij}^{\alpha \beta} \Sigma^\alpha \Gamma^\beta$, where $\tilde{\xi}^{\alpha \beta}$ = $\xi^{\alpha \beta}$ if $\alpha \in \{1,2\}$ and $\tilde{\xi}^{\alpha \beta}$ = $-\xi^{\alpha \beta}$ if $\alpha \in \{0,3\}$.

On the Kagome lattice, the allowed form of the gauge transformations has been determined by Yuan-Ming Lu {\it et al.}\cite{2011_lu} For completeness, we will reproduce that calculation, valid also for our spin triplet ansatz, here. The relations between the gauge transformation matrices,
\begin{gather}
[G_{T}(i)]^2 = \eta_{T} I, \label{eq_T}\\
G_{\sigma}(\sigma(i))G_{\sigma}(i) = \eta_{\sigma} I, \label{eq_sig}\\
G_{T_1}^\dag(i)G_{T}^\dag(i)G_{T_1}(i)G_{T}(T_1^{-1}(i)) = \eta_{T_1 T} I, \label{eq_1T}\\
G_{T_2}^\dag(i)G_{T}^\dag(i)G_{T_2}(i)G_{T}(T_2^{-1}(i)) = \eta_{T_2 T} I, \label{eq_2T}\\
G_{\sigma}^\dag(i)G_{T}^\dag(i)G_{\sigma}(i)G_{T}(\sigma^{-1}(i)) = \eta_{\sigma T} I, \label{eq_sigT}\\
G_{S_6}^\dag(i)G_{T}^\dag(i)G_{S_6}(i)G_{T}(S_6^{-1}(i)) = \eta_{S_6 T} I,\label{eq_ST}\\
G_{T_2}^\dag(T_1^{-1}(i))G_{T_1}^\dag(i)G_{T_2}(i)G_{T_1}(T_2^{-1}(i))=\eta_{12} I, \label{eq_12}\\
G_{S_6}(S_6^{-1}(i))G_{S_6}(S_6^{-2}(i))G_{S_6}(S_6^{3}(i))\nonumber\\
\times G_{S_6}(S_6^{2}(i))G_{S_6}(S_6(i))G_{S_6}(i) = \eta_{S_6} I,\label{eq_S}\\
G_{\sigma}^\dag(T_2^{-1}(i))G_{T_2}^\dag(i)G_{\sigma}(i)G_{T_1}(\sigma(i)) = \eta_{\sigma T_1} I,\label{eq_sig1}\\
G_{\sigma}^\dag(T_1^{-1}(i))G_{T_1}^\dag(i)G_{\sigma}(i)G_{T_2}(\sigma(i)) = \eta_{\sigma T_2} I,\label{eq_sig2}\\
G_{\sigma}^\dag(S_6(i))G_{S_6}(S_6(i))G_{\sigma}(i)G_{S_6}(\sigma(i)) = \eta_{\sigma S_6}I,\label{eq_sigS}\\
G_{S_6}^\dag(T_2^{-1}(i))G_{T_2}^\dag(i)G_{S_6}(i)G_{T_1}(S_6^{-1}(i))= \eta_{S_6 T_1}I,\label{eq_S1}\\
G_{S_6}^\dag(T_2^{-1}T_1(i))G_{T_2}^\dag(T_1(i))G_{T_1}(T_1(i))\nonumber\\
G_{S_6}(i)G_{T_2}(S_6^{-1}(i)) = \eta_{S_6 T_2}I\label{eq_S2},
\end{gather}
are valid for our case as well, due to the decoupling of spin and gauge portions of our ansatz. In the above, the relations are valid for all lattice sites $i=(x,y,s)$, I is the 4x4 identity matrix, and the $G_S$ matrices are gauge transformation matrices generated by exponentiation of the $\Gamma$ matrices. The $\eta$'s are $\pm$1, the choice of which characterize different spin liquid states. In deriving this form of the commutation relations, we have included a gauge transformation $i\Gamma^2$ in our definition of the time reversal operator, as this simplifies the effect of the operator on the mean field ansatz.

We turn next to the calculation of the gauge transformations. We look first at the gauge transformations associated with the translations. We can perform a site dependent gauge transformation $W(i)$, under which the gauge transformations associated with the translational symmetries transform as
\begin{align}
G_{T_1}(i) &\rightarrow W(i) G_{T_1}(i)W^\dag(i-\hat{x})\\
G_{T_2}(i) &\rightarrow W(i) G_{T_2}(i)W^\dag(i-\hat{y}).
\end{align}
As such, we can choose a gauge transformation W(i) to simplify the form of $G_{T_1}$ and $G_{T_2}$. Using such a transformation, along with condition \ref{eq_12}, we can restrict the form of these gauge transformations to be
\begin{align}
G_{T_1}(i) = \eta_{12}^{i_y}I \quad G_{T_2}(i) = I.
\end{align}
To preserve this choice, we can now only perform gauge transformations which are equivalent on all lattice positions ($W(x,y,s) = W(s)$) or transformations which change the shown matrices by an IGG transformation.

Next, we look at adding the reflection symmetry $\sigma$. Given our formulae for $G_{T_1}$ and $G_{T_2}$, along with the relations between the gauge transformations, we have that
\begin{align}
G_{\sigma}^\dag(T_2^{-1}(i))G_{\sigma}(i)\eta_{12}^x = \eta_{\sigma T_1}I\\
G_{\sigma}^\dag(T_1^{-1}(i))G_{\sigma}(i)\eta_{12}^y = \eta_{\sigma T_2}I.
\end{align}
Defining $G_{\sigma}$(0,0,s) = $g_{\sigma}$(s), we have, by repeated application of the above,
\begin{align}
G_{\sigma}(0,y,s)&=\eta_{\sigma T_1}^yg_{\sigma}(s)\\
G_{\sigma}(x,y,s)&=\eta_{\sigma T_1}^y\eta_{12}^{xy}\eta_{\sigma T_2}^xg_{\sigma}(s).
\end{align}
Next, using
\begin{align}
G_{\sigma}(\sigma(i))G_{\sigma}(i)= \eta_{\sigma}I
\end{align}
we find that
\begin{align}
\eta_{\sigma}I &= G_{\sigma}(y,x,\sigma(s))G_{\sigma}(x,y,s)\\
&= (\eta_{\sigma T_1}\eta_{\sigma T_2})^{x+y}g_{\sigma}(\sigma(s))g_{\sigma}(s).
\end{align}
Since this is true for all x and y, $\eta_{\sigma T_1}\eta_{\sigma T_2}=1$ and thus $\eta_{\sigma T_1}=\eta_{\sigma T_2}$ and $g_{\sigma}(\sigma(s))g_{\sigma}(s) = \eta_{\sigma}I$ (where $\sigma(u) = u, \sigma(v) = w$ and $\sigma(w) = v$). Our final form for the gauge transformation is
\begin{align}
G_{\sigma}(x,y,s) = \eta_{\sigma T_1}^{x+y}\eta_{12}^{xy}g_{\sigma}(s).
\end{align}

\begin{table*}
\caption{We list the solutions of Eq. \ref{eq_GT1} - \ref{eq_gST}, along with a set of gauge transformations which realize these solutions.}
\begin{center}
{\renewcommand{\arraystretch}{1.5}
\renewcommand{\tabcolsep}{0.2cm}
	\begin{tabular}{c|ccccccc|cccccc}
		\hline
		No. & $\eta_T$ & $\eta_{\sigma T}$ & $\eta_{S_6 T}$ & $\eta_\sigma$ & $\eta_{\sigma S_6}$ & $\eta_{S_6}$ & $\eta_{12}$ & $g_\sigma(u)$ & $g_\sigma(v)$ & $g_\sigma(w)$ & $g_{S_6}(u)$ & $g_{S_6}(v)$ & $g_{S_6}(w)$ \\ \hline
		1,2 & -1 & 1 & 1 & 1 & 1 & $\pm$1 & $\pm$1 & $\Gamma^0$ & $\Gamma^0$ & $\Gamma^0$ & $\Gamma^0$ & $\Gamma^0$ & $\Gamma^0$ \\ \hline
		3,4 & -1 & 1 & 1 & 1 & -1 & $\mp$1 & $\pm$1 & $\Gamma^0$ & $\Gamma^0$ & $\Gamma^0$ & $\Gamma^0$ & -$\Gamma^0$ & $i\Gamma^1$ \\ \hline
		5,6 & -1 & 1 & -1 & 1 & -1 & $\mp$1 & $\pm$1 & $\Gamma^0$ & $\Gamma^0$ & $\Gamma^0$ & $i\Gamma^3$ & $i\Gamma^3$ & $i\Gamma^3$ \\ \hline
		7,8 & -1 & 1 & 1 & -1 & -1 & $\mp$1 & $\pm$1 & $i\Gamma^1$ & $\Gamma^0$ & -$\Gamma^0$ & $\Gamma^0$ & $i\Gamma^1$ & $\Gamma^0$ \\ \hline
		9,10 & -1 & 1 & 1 & -1 & 1 & $\pm$1 & $\pm$1 & $i\Gamma^1$ & $\Gamma^0$ & -$\Gamma^0$  & $\Gamma^0$ & -$i\Gamma^1$ & $i\Gamma^1$ \\ \hline
		11,12 & -1 & 1 & -1 & -1 & 1 & $\mp$1 & $\pm$1 & $i\Gamma^1$ & $\Gamma^0$ & -$\Gamma^0$  & $i\Gamma^3$ & -$i\Gamma^2$ & $i\Gamma^3$ \\ \hline
		13,14 & -1 & -1 & -1 & -1 & -1 & $\mp$1 & $\pm$1 & $i\Gamma^3$ & $i\Gamma^3$ & $i\Gamma^3$ & $i\Gamma^3$ & $i\Gamma^3$ & $i\Gamma^3$ \\ \hline
		15,16 & -1 & -1 & 1 & -1 & 1 & $\pm$1 & $\pm$1 & $i\Gamma^3$ & $i\Gamma^3$ & $i\Gamma^3$ & $\Gamma^0$ & $\Gamma^0$ & $\Gamma^0$ \\ \hline
		17,18 & -1 & -1 & 1 & -1 & 1 & $\mp$1 & $\pm$1 & $i\Gamma^3$ & $i\Gamma^3$ & $i\Gamma^3$ & $\Gamma^0$ & $\Gamma^0$ & $i\Gamma^1$ \\ \hline
		19,20 & -1 & -1 & -1 & -1 & 1 & $\mp$1 & $\pm$1 & $i\Gamma^3$ & $i\Gamma^3$ & $i\Gamma^3$ & $i\Gamma^3$ & -$i\Gamma^3$ & $i\Gamma^3$ \\ \hline
		21,22 & 1 & 1 & 1 & 1 & 1 & $\pm$1 & $\pm$1 & $\Gamma^0$ & $\Gamma^0$ & $\Gamma^0$ & $\Gamma^0$ & $\Gamma^0$ & $\Gamma^0$ \\ \hline
		23,24 & 1 & 1 & 1 & 1 & -1 & $\mp$1 & $\pm$1 & $\Gamma^0$ & $\Gamma^0$ & $\Gamma^0$ & $\Gamma^0$ & -$\Gamma^0$ & $i\Gamma^3$ \\ \hline
		25,26 & 1 & 1 & 1 & -1 & -1 & $\mp$1 & $\pm$1 & $i\Gamma^3$ & $\Gamma^0$ & -$\Gamma^0$ & $\Gamma^0$ & $i\Gamma^3$ & $\Gamma^0$ \\ \hline
		27,28 & 1 & 1 & 1 & -1 & 1 & $\mp$1 & $\pm$1 & $i\Gamma^3$ & $\Gamma^0$ & -$\Gamma^0$ & $\Gamma^0$ & -$i\Gamma^3$ & $i\Gamma^1$ \\ \hline
		29,30 & 1 & 1 & 1 & -1 & 1 & $\pm$1 & $\pm$1 & $i\Gamma^3$ & $\Gamma^0$ & -$\Gamma^0$ & $\Gamma^0$ & -$i\Gamma^3$ & $i\Gamma^3$ \\ \hline
	\end{tabular}}
\end{center}
\label{tab_solns}
\end{table*}

Next we look at adding the $S_6$ symmetry to our calculation. We can do an IGG transformation, taking $G_{T_1}(T_1(i))$ to $\eta_{S_6T_2}G_{T_1}(T_1(i))$, with the net effect being that $\eta_{S_6T_2}$ becomes one (previous calculations are unaffected). We now have that
\begin{align}
G_{S_6}^\dag(T_2^{-1}T_1(i))G_{S_6}(i)\eta_{12}^y &= I \\
G_{S_6}^\dag(T_2^{-1}(i))G_{S_6}(i) \eta_{12}^{-x-1} &= \eta_{S_6 T_1} I \qquad (s=u,v)\\
G_{S_6}^\dag(T_2^{-1}(i))G_{S_6}(i) \eta_{12}^{-x} &= \eta_{S_6 T_1} I \qquad (s=w).
\end{align}
Defining $G_{S_6}(0,0,s) = g_{S_6}(s)$, we find that
\begin{align}
G_{S_6}(n,-n,s) &= \eta_{12}^{n(n-1)/2}g_{S_6}(s)\\
G_{S_6}(x,y,s) &= \eta_{12}^{x(x-1)/2 +y +xy}\eta_{S_6 T_1}^{x+y} g_{S_6}(s) \qquad (s=u,v) \\
G_{S_6}(x,y,s) &= \eta_{12}^{x(x-1)/2 +xy}\eta_{S_6 T_1}^{x+y} g_{S_6}(s) \qquad (s=w).
\end{align}
Using the commutation relation between the $\sigma$ and $S_6$ gauge transformations, we find that
\begin{align}
\eta_{\sigma S_6} I &= \eta_{\sigma T_1}^y\eta_{12}^y \eta_{S_6 T_1}^y g_{\sigma}^\dag(v)g_{S_6}(v)g_{\sigma}(u)g_{S_6}(u) \\
&= \eta_{\sigma T_1}^y\eta_{12}^y \eta_{S_6 T_1}^y \eta_{\sigma} g_{\sigma}(w)g_{S_6}(v)g_{\sigma}(u)g_{S_6}(u)
\end{align}
giving us that $\eta_{\sigma T_1}\eta_{12}\eta_{S_6 T_1} = 1$ and $g_{\sigma}(u)g_{S_6}(u)g_{\sigma}(w)g_{S_6}(v) = \eta_{\sigma S_6}\eta_{\sigma} I$. A similar calculation on a different sublattice gives us
\begin{align}
\eta_{\sigma S_6} I &= \eta_{\sigma T_1}^y\eta_{12}^y \eta_{S_6 T_1}^y g_{\sigma}^\dag(w)g_{S_6}(w)g_{\sigma}(v)g_{S_6}(w) \\
&= \eta_{\sigma T_1}^y\eta_{12}^y \eta_{S_6 T_1}^y \eta_{\sigma} g_{\sigma}(v)g_{S_6}(w)g_{\sigma}(v)g_{S_6}(w)
\end{align}
giving us $(g_{\sigma}(v)g_{S_6}(w))^2 = \eta_{\sigma S_6}\eta_{\sigma} I$. A $Z_2$ (IGG) gauge transformation of the form $W(x,y,s) = \eta_{\sigma T_1}^y$ changes $\eta_{\sigma T_1}$ to 1. Using the cyclic relation of the gauge transformations related to the $S_6$ operators, we find
\begin{align}
\eta_{S_6} I = \eta_{12} (g_{S_6}(w)g_{S_6}(v)g_{S_6}(u))^2
\end{align}
giving us that
\begin{align}
[g_{S_6}(w)g_{S_6}(v)g_{S_6}(u)]^2 = \eta_{S_6} \eta_{12} I.
\end{align}

Next we turn to the time reversal symmetry. Similar methods to the above give us that
\begin{align}
[G_{T}(i)]^2 &= \eta_{T}I \\
G_{T}^\dag(i)G_{T}(i+\hat{x}) &= \eta_{T_1 T} I\\
G_{T}^\dag(i)G_{T}(i+\hat{y}) &= \eta_{T_2 T} I.
\end{align}
The first of these relations tells us that $G_{T}(i)$ is either the identity (for $\eta_{T} =1$) or $i\vec{a}\cdot\vec{\kappa}$ (for $\eta_{T}=-1$, where $|\vec{a}|=1$. Defining $G_{T}(0,0,s) = g_{T}(s)$,
\begin{align}
G_{T}(x,y,s) = \eta_{T_1 T}^x \eta_{T_2 T}^y g_{T}(s)
\end{align}
and further, using the commutation relations between the $\sigma$ and $T$ gauge transformations and the $S_6$ and $T$ gauge transformations,
\begin{align}
g_{\sigma}^\dag(s) g_{T}^\dag(s) g_{\sigma}(s)g_{T}(\sigma(s)) \eta_{T_1 T}^{x+y} \eta_{T_2 T}^{x+y} &= \eta_{\sigma T}I \\
g_{S_6}^\dag(s) g_{T}^\dag(s) g_{S_6}(s)g_{T}(S_6^{-1}(s)) \eta_{T_1 T}^{f_1(i)} \eta_{T_2 T}^{f_2(i)} &= \eta_{S_6 T}I.
\end{align}
Because this is true for all x and y, and $f_1(i)$ is not equal to $f_2(i)$, $\eta_{T_1 T} = \eta_{T_2 T} = 1$. If $G_{T}(i) = i\vec{a}\cdot \vec{\kappa}$, we perform a gauge transformation W on $G_{T}(i)$ such that $W^\dag G_{T}(i) W = i\kappa^1$ (as this is the same on all sites, it does not affect our gauge fixing for the translation gauge transformations). Collecting the necessary results for further use,
\begin{gather}
G_{T_1}(x,y,s) = \eta_{12}^y I \label{eq_GT1}\\
G_{T_2}(x,y,s) = I \label{eq_GT2}\\
G_{\sigma}(x,y,s) = \eta_{12}^{xy}g_\sigma(s) \label{eq_Gsig}\\
G_{S_6}(x,y,s) = \eta_{12}^{xy+(x+1)x/2}g_{S_6}(s) \qquad s=u,v \label{eq_GS6uv}\\
G_{S_6}(x,y,s) = \eta_{12}^{xy+x+y+(x+1)x/2}g_{S_6}(s) \qquad s=w \label{eq_GS6w}\\
G_{T}(s) = I = g_{T}(s) \qquad \eta_{T} = 1 \label{eq_GTI}\\
G_{T}(s) = i\Gamma^1 = g_{T}(s) \qquad \eta_{T} = -1 \label{eq_GTN}\\
g_{\sigma}(\sigma(s))g_{\sigma}(s) = \eta_{\sigma}I \label{eq_gsigsig}\\
g_{\sigma}(u)g_{S_6}(u)g_{\sigma}(w)g_{S_6}(v) = (g_{\sigma}(v)g_{S_6}(w))^2 = \eta_{\sigma S_6}\eta_{\sigma} I \label{eq_gsigSsigS}\\
(g_{S_6}(w)g_{S_6}(v)g_{S_6}(u))^2 = \eta_{S_6}\eta_{12}I \label{eq_gSSS}\\
g_{\sigma}(s)g_{T}(\sigma(s)) = \eta_{\sigma T}g_{T}(s)g_{\sigma}(s) \label{eq_gsigT}\\
g_{S_6}(s)g_{T}(S_6^{-1}(s)) = \eta_{S_6 T}g_{T}(s)g_{S_6}(s). \label{eq_gST}
\end{gather}
We also have the gauge freedom left to perform a gauge rotation arbitrarily at all positions for $\eta_{T} = 1$ or an arbitrary gauge rotation about the x axis for $\eta_{T} = -1$.

The solution to the above equations is derived in detail by Lu {\it et al.}\cite{2011_lu} and as such we simply list the results in table \ref{tab_solns}. The basic method of obtaining these solutions is as follows: for each choice of $Z_2$ parameter set, we determine whether there is a choice of gauge matrices \{ $g_S$ \} which satisfy the equations \ref{eq_GT1} - \ref{eq_gST}. In order to do so, we determine the allowed forms of the $g_S$ matrices from the equations, then use the gauge freedom on each site to fix the form of these. Of particular not is the fact that in the consistency equations for the $g$ matrices, the terms $\eta_{12}$ and $\eta_{S_6}$ only appear multiplied together, meaning that for any choice of the gauge matrices $g_S$ we can choose $\eta_{12} = \pm 1$, which fixes the form of $\eta_{S_6}$.

\section{Relation among the mean-field paramters}
\label{appen_param_relation}
The relation among the different singlet and triplet parameters in terms of $\xi_{ij}$ is given by
\begin{gather}
\label{eq_rels}
\chi_{ij}=\xi_{ij}^{00}+\xi_{ij}^{03};~~~~~\eta_{ij}=-\xi_{ij}^{01}+i\xi_{ij}^{02};\nonumber\\
E^1_{ij}=\xi_{ij}^{10}+\xi_{ij}^{13};~~~~E^2_{ij}=\xi_{ij}^{20}+\xi_{ij}^{23};~~~E^3_{ij}=\xi_{ij}^{30}+\xi_{ij}^{33}\nonumber\\
D^1_{ij}=-\xi_{ij}^{11}+i\xi_{ij}^{12};~~D^2_{ij}=-\xi_{ij}^{21}+i\xi_{ij}^{22};~~D^3_{ij}=-\xi_{ij}^{31}+i\xi_{ij}^{32}
\end{gather}

Using these, we can derive the form of the bond nematic order parameter and vector chirality order parameters, which are given in terms of the mean field parameters\cite{2009_shindou} as
\begin{align}
\mathcal{Q}_{ij}^{\mu,\nu} =&-\frac{1}{2}\big( E_{ij}^{\mu} E_{ij}^{\ast \nu} - \frac{1}{3} \delta^{\mu,\nu}|\vec{E}_{ij}|^2\big) + h.c.\nonumber\\
&-\frac{1}{2}\big( D_{ij}^\mu D_{ij}^{\ast \nu} - \frac{1}{3} \delta^{\mu,\nu}|\vec{D}_{ij}|^2\big) + h.c.\nonumber\\
\mathcal{J}_{ij}^\lambda =& \frac{i}{2} \big( \chi_{ij} E_{ij}^{\ast \lambda} - \chi_{ij}^\ast E_{ij}^{\lambda}\big)\nonumber \\
&+ \frac{i}{2} \big( \eta_{ij} D_{ij}^{\ast \lambda} - \eta_{ij}^\ast D_{ij}^{\lambda}\big)
\end{align}
where our definition of $\eta_{ij}$ differs by a factor of (-1) from that of the cited work. We rewrite this in terms of our variables, finding
\begin{align}
\mathcal{Q}_{ij}^{\mu \nu} =& -\xi_{ij}^{\mu 0}\xi_{ij}^{\nu 0} + \sum_a \xi_{ij}^{\mu a}\xi_{ij}^{\nu a} \nonumber\\
+& \frac{\delta^{\mu \nu}}{3}\sum_{b}\big( (\xi_{ij}^{b 0})^2 - \sum_{a}(\xi_{ij}^{ba})^2)\nonumber\\
\mathcal{J}_{ij}^\lambda =& i(\xi_{ij}^{00}\xi_{ij}^{\lambda 0} - \sum_{a} \xi_{ij}^{0a}\xi_{ij}^{\lambda a} )
\end{align}
In particular, we find that $\mathcal{J}^1$, $\mathcal{J}^2$, $\mathcal{Q}^{13}$ and $\mathcal{Q}^{23}$ must be zero for all non-Kramers spin liquids, as the terms allowed by symmetry in Eq. \ref{eq_eta1} and \ref{eq_eta-1} do not allow non-zero values for these order parameters. 



\begin{thebibliography}{30}
\bibitem{1999_fazekas} P. Fazekas, {\it Lecture Notes on Electron Correlation and Magnetism (Series in Modern Condensed Matter Physics)}, World Scientific Pub Co Inc (1999).

\bibitem{2001_yosida} K. Yosida, {\it Theory of Magnetism}, Springer (2001).

\bibitem{1952_bleany} B. Bleaney and H. E. D. Scovil, Phil. Mag. 43, 999 (1952).

\bibitem{2011_onada} Shigeki Onoda and Yoichi Tanaka, \href{http://prb.aps.org/abstract/PRB/v83/i9/e094411}{\prb 83, 094411(R) (2011).}

\bibitem{2002_matsuhira} K. Matsuhira, Y. Hinatsu, K. Tenya1, H. Amitsuka and T. Sakakibara, \href{http://jpsj.ipap.jp/link?JPSJ/71/1576/}{J. Phys. Soc. Jpn. 71, 1576 (2002).}

\bibitem{2009_matsuhira} K. Matsuhira, C. Sekine, C. Paulsen, M. Wakeshima, Y. Hinatsu, T. Kitazawa, Y. Kiuchi, Z. Hiroi, and S. Takagi, \href{http://iopscience.iop.org/1742-6596/145/1/012031/}{J. Phys. Conf. Ser. 145, 012031 (2009)}.

\bibitem{2006_nakatsuji} S. Nakatsuji, Y. Machida, Y. Maeno, T. Tayama, T. Sakakibara, J. van Duijn, L. Balicas, J. N. Millican, R. T. Macaluso, and J. Y. Chan, \href{http://prl.aps.org/abstract/PRL/v96/i8/e087204}{Phys. Rev. Lett. 96, 087204 (2006).}

\bibitem{2012_sungbin} SB. Lee, S. Onoda, and L. Balents, \href{http://prb.aps.org/abstract/PRB/v86/i10/e104412}{Phys. Rev. B 86, 104412 (2012).}

\bibitem{2013_flint} R. Flint, and T. Senthil, \href{http://arxiv.org/abs/1301.0815}{arXiv:1301.0815 (2013).}

\bibitem{2013_chandra} P. Chandra, P. Coleman, and R. Flint, \href{http://www.nature.com/nature/journal/v493/n7434/full/nature11820.html?WT.ec_id=NATURE-20130131}{Nature 493, 621 (2013).}

\bibitem{2006_suzuki} O. Suzuki, H. S. Suzuki, H. Kitazawa, G. Kido, T. Ueno, T. Yamaguchi, Y. Nemoto, T. Goto, J. Phys. Soc. Japan 75, 013704 (2006).

\bibitem{1930_kramers} H. A. Kramers, Proc. Amsterdam Acad. 33, 959 (1930).

\bibitem{2002_wen} X. G. Wen, \href{http://prb.aps.org/abstract/PRB/v65/i16/e165113}{\prb 65, 165113 (2002)}; X.-G. Wen, Quantum Field Theory of Many-Body Systems. Oxford University Press, Oxford (2004).

\bibitem{1987_anderson}P. W. Anderson, G. Baskaran, Z. Zou, and T. Hsu, \href{http://prl.aps.org/abstract/PRL/v58/i26/p2790_1}{\prl 58, 2790 (1987).}

\bibitem{2011_lu} Yuan-Ming Lu, Ying Ran, and Patrick A. Lee, \href{http://http://prb.aps.org/abstract/PRB/v83/i22/e224413}{\prb 83, 224412 (2011).}

\bibitem{2013_dodds}T. Dodds, S. Bhattacharjee, and Y. B. Kim, \href{http://arxiv.org/abs/1303.1154}{arXiv:1303.1154}

\bibitem{2012_schaffer} R. Schaffer, S. Bhattacharjee, and Y. B. Kim, \href{Phys. Rev. B 86, 224417 (2012)}{\prb 86, 224417 (2012).}

\bibitem{1988_affleck} I. Affleck, and J. B. Marston, \href{http://prb.aps.org/abstract/PRB/v37/i7/p3774_1}{\prb 37, 3774 (1988).}

\bibitem{1992_lee} P. A. Lee, and N. Nagaosa, \href{http://prb.aps.org/abstract/PRB/v46/i9/p5621_1}{\prb 46, 5621 (1992).}

\bibitem{2013_lee} P. A. Lee, and N. Nagaosa, \href{http://prb.aps.org/abstract/PRB/v87/i6/e064423}{\prb 87, 064423 (2013).}

\bibitem{2009_shindou} R. Shindou, and T. Momoi, \href{http://prb.aps.org/abstract/PRB/v80/i6/e064410}{\prb 80, 064410 (2009).}

\bibitem{2012_bhattacharjee} S. Bhattacharjee, Y. B. Kim, S.-S. Lee, and D.-H. Lee, \href{http://prb.aps.org/abstract/PRB/v85/i22/e224428}{\prb 85, 224428 (2012).}

\end{thebibliography}
\end{document}